# Max-Cut and Max-Bisection are NP-hard on unit disk graphs[*]

Josep Díaz[1] and Marcin Kamiński[2]

[1]Llenguatges i Sistemes Informàtics

Universitat Politècnica de Catalunya

08034 Barcelona

diaz@lsi.upc.edu

[2] RUTCOR, Rutgers University

640 Bartholomew Road

Piscataway, NJ 08854, USA

mkaminski@rutcor.rutgers.edu

June 28, 2018

## Abstract

We prove that the MAX-CUT and MAX-BISECTION problems are NP-hard on unit disk graphs. We also show that $\lambda$-precision graphs are planar for $\lambda > 1/\sqrt{2}$.

## 1 Introduction

Given a simple graph $G$, *cut* in $G$ is a partition of its vertex set into two parts. A cut whose parts have the same cardinality is called a *bisection*. The size of a cut is the number of edges that have their endpoints into two different parts of the cut. A *maximum cut* is a cut with maximum size and the algorithmic MAX-CUT problem consists in finding one given the input graph. Similarly, a *maximum bisection* is a a bisection with maximum size and the MAX-BISECTION problem is to find one. Let $mc(G)$ denote the maximum size of a cut in $G$.

The MAX-CUT problem is known to be NP-hard for general graphs ([11]) and remains NP-hard even when the input graph is restricted to be a split or 3-colorable graph ([3]). As shown in [14], the problem is NP-hard also for the class of graphs with bounded maximum degree $\Delta$, if $\Delta \geq 3$. On the other hand, the MAX-CUT problem can be solved in polynomial time for planar graphs ([8]) or graphs with bounded treewidth ([3]).

The MAX-BISECTION problem is NP-hard for general graphs [6]. However, contrary to the MAX-CUT problem, MAX-BISECTION remains NP-hard on planar graphs (result of Jerrum presented in [10]).

[*]Research was partially supported by the EC Research Training Network HPRN-CT-2002-00278 (COMBSTRU) and the Spanish CYCIT: TIN2004-07925-C03-01 (GRAMMARS). The first author was also supported by *La distincio per a la promació de la recerca de la Generalitat de Catalunya*, 2002. The work was done when the second author was visiting Universitat Politècnica de Catalunya.

Unit disk graphs are intersection graphs of unit diameter disks in the plane. Place $n$ disks of diameter one in the plane so that the centers of disks do not coincide. An undirected graph is said to be a *unit disk graph* if there exists a one-to-one correspondence between its vertices and disks in such a way that two vertices are adjacent if and only if the corresponding disks intersect. (We assume tangent disks do intersect, however the classes of unit disk graphs with open or closed disks coincide [12].) Each configuration of disks that defines a unit disk graph is called its *intersection model*. This can be easily translated into a *proximity model* which is a collection of distinct points on the plane in one-to-one correspondence to the vertices of the graph in such a way that two vertices are adjacent if and only if two points are at distance at most one. Notice that unit disk graphs are simple and loopless. Often it is convenient to identify points in the proximity model and vertices of the graph and we will often do so.

In recent years there has been an increasing interest in the study of unit discs graphs and their randomized version, the random geometric graphs, due to their use as models of wireless communication networks (see [1]).

Sometimes an additional assumption on the proximity model is imposed and points are required to be at distance at least $\lambda$ from each other. Graphs which have such a proximity model are called $\lambda$-precision unit disk graphs. Notice that the class of $\lambda_1$-precision unit disk graphs is contained in the class of $\lambda_2$-precision unit disk graphs, for every $\lambda_1 \geq \lambda_2$ [9].

In [4], many algorithmic problems – CHROMATIC NUMBER, MAXIMUM INDEPENDENT SET, MINIMUM DOMINATING SET – have been proved to be NP-hard on unit disk graphs. The authors of that paper also provide a polynomial-time algorithm for MAX-CLIQUE. They conclude the paper pointing that computational complexities of every studied problem agree on planar and unit disk graphs.

The computational complexity of the MAX-CUT and MAX-BISECTION problem on unit disk graphs has not been known, even though several authors offered approximation algorithms for MAX-CUT and MAX-BISECTION. In [9], among other approximation results, the authors present a polynomial-time approximation scheme for $\lambda$-precision unit disk graphs. A polynomial-time approximation scheme for the MAX-BISECTION problem on unit disk graphs was developed in [10]. It is also worth mentioning that the problem of recognizing unit disc graphs is NP-hard ([2]) so the approximation algorithms require providing an intersection model for the input graph.

In this paper, we prove that the MAX-CUT and the MAX-BISECTION problems are NP-hard for unit discs graphs, which to the knowledge of the authors were open problems (see [10, 13]). Also, it turns out that the MAX-CUT is the first problem known whose computational complexities on planar and unit disk graphs do not agree. In the last section, we show that the $\lambda$-precision unit disk graphs are planar for $\lambda > 1/\sqrt{2}$. An interesting open problem is to investigate the computational complexity of MIN-BISECTION (i.e. the problem of finding a bisection of minimum size) on unit disk graphs ([5]).

## 2 Mesh drawings

A *drawing of a graph* $G = (V, E)$ is a mapping $f$ which assigns to each vertex of $G$ a distinct point in the plane and to each edge $uv$ a continuous arc between $f(u)$ and $f(v)$, not passing through the image of any other vertex. We also allow interiors of images of two different edges to intersect only at a finite number of points. Each such intersection is called a *crossing point*. A graph which can be drawn in the plane without any crossing points is called a *planar graph*. Below, if it does not lead to misunderstanding, we often do not distinguish between vertex/edge and its image.

The *mesh* $\mathcal{M}$ is the set of points in the plane which have at least one integral coordinate. These points of the mesh that have two integral coordinates are called *mesh crosses*. The $x$-distance ($y$-distance) between two points of $\mathcal{M}$ is the difference of their $x$ ($y$) coordinates.

A *mesh drawing* of a graph is a drawing in which the images of all vertices are mesh crosses and the images of edges belong to the mesh. Notice that in a mesh drawing only two edges can intersect at a crossing point and a crossing point always occurs at a mesh cross.

A necessary condition for a graph to have a mesh drawing is to be of maximum degree 4. Now we will show that in fact that condition is also sufficient.

**Lemma 1.** *Every graph of maximum degree 4 has a mesh drawing and the drawing can be found in polynomial time.*

*Proof.* Let $G$ be any graph of maximum degree 4. Place all vertices of $G$ at distinct mesh crosses in such a way that the $x$- and $y$-distance between any pair of vertices is at least 5. For a vertex of $G$ whose corresponding mesh cross has coordinates $(a, b)$ we define four "corridors" which are the sets of points described by the following equations:

A. $x = a - 1$ or $y = b + 2$,

B. $x = a - 2$ or $y = b + 1$,

C. $x = a + 1$ or $y = b - 2$

D. $x = a + 2$ or $y = b - 1$.

Then an edge between any pair of vertices in $\mathcal{M}$ can be drawn entirely within the corridors of its endpoints. Hence we can construct a mesh drawing of $G$ and the construction is can be done in polynomial time. □

For the reduction described below we need a mesh drawing in which distances between objects (vertices, edges and crossing points) are large enough. A mesh drawing is called *standard* if

(i) the distance between any two crossing points is at least 10,

(ii) the distance between any two vertices is at least 10,

(iii) the distance between any vertex and any crossing point is at least 10,

(iv) if interior points of two edges belong to two different parallel vertical ($x = k$ for some $k \in \mathbb{Z}$) or horizontal ($y = k$ for some $k \in \mathbb{Z}$) lines, the distance between these two lines is at least 10.

Let us show how, given a mesh drawing of a graph, we can create its standard mesh drawing. Suppose that the distance between two vertices $v, w$ is less than 10. Pick a line which separates these two vertices (i.e. at most one of $v, w$ belongs to the line). Without loss of generality, we can assume it is a horizontal line $y = k$ for some $k \in \mathbb{Z}$ and $v$ lies above the line. Now all the points of the drawing whose $y$-coordinate is at most $k$ will be moved down by some constant $c$. The constant should be such that after $w$ is moved down by $c$ the distance between $v$ and $w$ is at least 10. All the edges that were broken by that operation should be extended (by adding a vertical segment of length $c$) in a way that makes a new graph isomorphic to the original one. This way we decreased the number of pairs of vertices that were at distance at most 10 from each other. A similar technique can be used to satisfy all the conditions of the standard mesh drawing and a standard mesh drawing can be found in polynomial time.

**Lemma 2.** *Every graph of maximum degree 4 has a standard mesh drawing and the drawing can be found in time polynomial in the number of vertices of the graph.*

## 3 Gadget

In this section we are going to describe a gadget that will be used later for the reduction. However, first we need to prove a useful lemma.

**Lemma 3.** *Let $G'$ be the graph obtained from graph $G$ by subdividing one of the edges of $G$ twice. Then, $mc(G') = mc(G) + 2$.*

*Proof.* Let $uv$ be an edge of $G$ which was subdivided twice and let $a$, $b$ be the new vertices. Notice that in any maximum cut of $G'$, (i) each of the edges $va$, $bu$ is always in the cut and (ii) $ab$ is in the cut if and only if $vu$ is in the maximum cut in $G$. Now it is easy to see that $mc(G') = mc(G) + 2$. □

**Construction 4.** *Let $H$ be a graph on 8 vertices: $v_0, v_1, v_2, v_3, w_0, w_1, w_2, w_3$ such that*

(i) *$H[v_0, v_1, v_2, v_3]$ is a $K_4$,*

(ii) *$H[w_i, v_i, v_{i+1}]$ is a $K_3$, for $i = 0, 1, 2, 3$, where $i + 1$ is taken modulo 4,*

(iii) *$H$ has no other edges than those needed to satisfy (i) and (ii).*

**Lemma 5.** *$H$ is a $(1/\sqrt{2})$-precision unit disk graph.*

*Proof.* Let us consider the following proximity model for $H$. Place vertices $v_0, v_1, v_2, v_3$ at points $(1/2, 0)$, $(0, 1/2)$, $(-1/2, 0)$ and $(0, -1/2)$, respectively. Vertices $w_0, w_1, w_2, w_3$ should be put at points $(4/5, 4/5)$, $(-4/5, 4/5)$, $(-4/5, -4/5)$ and $(4/5, -4/5)$, respectively. Now it is an exercise to verify that the distance between a pair of vertices is at most 1 if and only if they are adjacent. Also notice that the distance between any two points is at least $1/\sqrt{2}$. Hence, $H$ is a $(1/\sqrt{2})$-precision unit disk graph. □

**Construction 6.** *Let $G$ be a graph and $v_0v_2$, $v_1v_3$ two of its edges not incident to each other. Consider a graph $G^*$ which is obtained from $G$ by adding new vertices $w_0, w_1, w_2, w_3$ and edges necessary to make $G^*[v_0, v_1, v_2, v_3, w_0, w_1, w_2, w_3]$ isomorphic to $H$. We say that $H$ was constructed in $G$ on edges $v_0v_2$, $v_1v_3$.*

**Lemma 7.** *Let $G^*$ be a graph obtained from $G$ by constructing $H$ on two non-incident edges of $G$. Then, $mc(G^*) = mc(G) + 8$*

*Proof.* Notice that the contribution of any of the triangles $w_i, v_i, v_{i+1}$ in the copy of $H$ ($i = 0, 1, 2, 3$ and $i + 1$ is taken modulo 4) to the maximum cut in $G^*$ is 2. Each edge of $H$ is either in one of the triangles (there are four of them) or is an original edge of $G$. Hence, $mc(G^*) = mc(G) + 8$. □

## 4 Reduction

**Theorem 8.** MAX-CUT *is NP-hard on unit disk graphs.*

*Proof.* To prove that MAX-CUT is NP-hard on unit disk graphs we are going to present a polynomial-time procedure that takes an arbitrary graph $G$ of maximum degree 4 and produces a unit disk graph $G'$. Moreover, knowing $mc(G')$ we are able compute $mc(G)$ in polynomial time.

STEP 1 Let $G$ be a graph of maximum degree 3. Let us consider its standard mesh drawing. According to Lemma 2, it does exist and can be found in polynomial time.

STEP 2 Subdivide edges of $G$ putting new vertices at mesh crosses that are not crossing points. Now we have two types of vertices – those that were created in this step and those that correspond to original vertices of $G$.

STEP 3 For each crossing point $(x, y)$, subdivide only the edge between $(x, y-1)$ and $(x, y+1)$ (the vertical one). We remove two vertices placed at $(x-1, y)$ and $(x+1, y)$ and consequently all the edges they were incident to. We place new vertices at coordinates $(x-1.5, y+0.5)$, $(x-0.5, y+0.5)$, $(x+0.5, y+0.5)$, $(x+1.5, y+0.5)$ and create edges between pairs of vertices: $(x-2, y)$ and $(x-1.5, y+0.5)$, $(x-1.5, y+0.5)$ and $(x-0.5, y+0.5)$, $(x-0.5, y+0.5)$ and $(x+0.5, y+0.5)$, $(x+0.5, y+0.5)$ and $(x+1.5, y+0.5)$, $(x+1.5, y+0.5)$ and $(x+2, y)$. Notice that the drawing is not a mesh drawing anymore.

STEP 4 For each crossing point, construct a copy of graph $H$ on the crossing edges. Place new vertices in the way described in the proof of Fact 5 and create straight line edges in each copy of $H$.

STEP 5 If along one of the original edges of $G$ there is an odd number of vertices, we need to subdivide one of the new edges once more. Pick an edge between points $(x, y)$ and $(x+1, y)$ whose both endpoints and the neighbors of the endpoints are of degree at most 2 and the endpoints belong to the original edge of $G$. Move vertex at $(x, y)$ to a new position $(x - 1/4, y)$ and vertex at $(x+1, y)$ to $(x + 5/4, y)$. Create a new vertex at $(x + 1/2, y + 1/2)$ and edges between pairs of vertices: $(x - 1/4, y)$ and $(x + 1/2, y)$, $(x + 5/4, y)$ and $(x + 1/2, y)$.

Let $U(G)$ be the graph whose drawing was constructed above. We will show that the graph is in fact a unit disk graph.

**Claim 9.** $U(G)$ is a $(1/\sqrt{2})$-precision unit disk graph.

*Proof of Claim.* To prove that $U(G)$ is a unit disk graph we will show that placing vertices in the plane at the same coordinates as in the construction above gives a a proximity model of $U(G)$.

Notice that after Step 2 all adjacent vertices, expect the endpoints of crossing edges, are at distance exactly 1. Once crossing edges have been replaced by a construction described in Step 3, all the neighbors of a vertex are within distance 1 from it. However, the distance between endpoints belonging to two crossing edges is $1/\sqrt{2}$, while these vertices are not adjacent.

Notice that after constructing a copy of $H$ at a crossing point (Step 4), "$w$" vertices are farther than a unit distance from all the vertices that they are not adjacent to. Also, the new edges created between endpoints belonging to two crossing edges are now connected. From this and Fact 5 follows that the graph obtained after Step 4 is a unit disk graph.

Observe that an edge whose both endpoints and the neighbors of the endpoints are of degree at most 2 can always be found if we start with a standard drawing. It is easy to see, that the construction described in Step 5 leaves a unit disk graph. □

Let $k$ be the number of crossing points in the standard mesh drawing of $G$ from Step 1 and $t$ the total number of subdivisions of all the original edges of $G$ after Step 5. Notice that $t$ must be even.

**Claim 10.** $mc(U(G)) = mc(G) + 8k + t$

*Proof of Claim.* A copy of $H$ is constructed on each pair of crossing edges and each copy of $H$ increases the value of maximum cut by 8 (Fact 7). Also, each double subdivision of an edge increases the value of maximum cut by 2 (Lemma 3). Hence, $mc(U(G)) = mc(G) + 8k + t$. □

We have shown that any graph $G$ of maximum degree 4 can be transformed in a unit disk graph $G'$ (Claim 9). If there exist a polynomial-time algorithm solving MAX-CUT on unit disk graphs, then knowing the construction and $mc(G')$, the value of maximum cut of $G'$ can be also computed in polynomial time. However, as shown in [14], MAX-CUT is NP-hard on graphs with maximum degree 4 and therefore it is also NP-hard on unit disk graphs. □

Taking two disjoint copies of a unit disk graph $G$ creates a unit disk graph whose maximum bisection is twice the value of maximum cut of $G$. The following fact is a simple corollary of Theorem 8.

**Corollary 11.** MAX-BISECTION *is NP-hard on unit disk graphs.*

## 5 Precision and planarity

Recall, that a unit disk graph is called $\lambda$-*precision* if the distance between any pair of vertices is at least $\lambda$. In this section we study the relation between $\lambda$-precision and planarity.

**Theorem 12.** *A $\lambda$-precision unit disk graph is planar, for every $\lambda > 1/\sqrt{2}$.*

*Proof.* Let $G$ be a $\lambda$-precision unit disk graph, for some $\lambda > 1/\sqrt{2}$. Consider a drawing of $G$ defined by its proximity model: put vertices in the plane at the same positions as in the proximity model and connect two of them by a straight line segment if and only if the distance between them is at most 1.

Let us consider an edge $e$ of $G$ of length $x$. The set of points at distance more than $1/\sqrt{2}$ from both endpoints of $e$ consists of two disjoint regions; one on each side of the straight line containing $e$. It is easy to verify that the distance between these two regions is strictly greater than $2\sqrt{1/2 - (x/2)^2}$ and therefore always strictly greater than 1. Hence, there is no edge crossing $e$ and the planarity of $G$ follows. □

**Theorem 13.** *The* MAX-CUT *problem is NP-hard in the class of $\lambda$-precision unit disk graphs if $\lambda \leq 1/\sqrt{2}$ and can be solved in polynomial time if $\lambda > 1/\sqrt{2}$*

*Proof.* Notice that the unit disk graph constructed in the proof of Theorem 8 is $(1/\sqrt{2})$-precision unit disk graph (Claim 9). Hence, MAX-CUT is NP-hard in the class of $\lambda$-precision unit disk graphs if $\lambda \leq 1/\sqrt{2}$.

MAX-CUT can be solved in polynomial time on planar graphs ([8]). Since $\lambda$-precision unit disk graphs are planar for $\lambda > 1/\sqrt{2}$ (Theorem 12), the MAX-CUT problem can be solved on these graphs in polynomial time. □